\documentstyle[12pt,epsfig]{article}
\begin{document}

\begin{titlepage}
\rightline{July 2014}
\vskip 2.2cm
\centerline{\Large \bf  
Can dark matter - electron scattering 
}
\vskip 0.4cm
\centerline{\Large \bf  
explain the DAMA annual modulation signal?}
\vskip 2.0cm
\centerline{R. Foot\footnote{
E-mail address: rfoot@unimelb.edu.au}}
\vskip 0.7cm
\centerline{\it ARC Centre of Excellence for Particle Physics at the Terascale,}
\centerline{\it School of Physics, University of Melbourne,}
\centerline{\it Victoria 3010 Australia}
\vskip 1.8cm
\noindent
% abstract here
The annually modulating $\sim$ keV scintillations observed in the DAMA/NaI and DAMA/Libra experiments 
might be due to dark matter - electron scattering. Such
an explanation is now favoured given the stringent constraints on nuclear recoil rates obtained by LUX, SuperCDMS
and other experiments. We suggest that multi-component dark matter models featuring light dark matter particles of
mass $\sim$ MeV can potentially explain the data. A specific example, kinetically mixed mirror dark matter, is shown
to have the right broad properties to consistently explain the experiments via dark matter - electron scattering.
If this is the explanation of the annual modulation signal found in the DAMA experiments
then a sidereal diurnal modulation signal is also anticipated.
We point out that the data from the DAMA experiments 
show a diurnal variation at around 2.3$\sigma$ C.L. with phase consistent with that expected.
This electron scattering interpretation of the DAMA experiments can potentially be probed in large
xenon experiments (LUX, XENON1T,...), as well as in low threshold experiments (CoGeNT, CDEX, C4, ...)
by searching for annually and diurnally modulated electron recoils.  

\end{titlepage}

\section{Introduction}

%\vskip 0.4cm

The DAMA/NaI and DAMA/Libra experiments \cite{dama1, dama2}
have provided very convincing evidence for dark matter direct detection.
These experiments have observed an annual modulation in the `single hit' event rate,
at around $9\sigma$ C.L., 
with a period and phase consistent with dark matter interactions \cite{spergel} in a NaI detector.
The DAMA experiments are sensitive to dark matter (DM) scattering off atomic electrons and/or nuclei.
An explanation in terms of nuclear recoils appears to be unlikely in view of the null results
reported by LUX \cite{lux}, XENON100 \cite{xenon}, SuperCDMS \cite{cdms}, CRESST-II \cite{cresst} and other experiments.
These experiments have provided very sensitive constraints on the rate of dark matter scattering
off nuclei, but can be relatively insensitive to dark matter scattering off atomic electrons.
An explanation of DAMA in terms of electron scattering thus seems to be favored if one hopes to 
explain the DAMA results consistently with the null results of other experiments.

Observable (keV) energy depositions in dark matter detectors can be produced via
DM - electron scattering if the Milky Way dark matter halo
contains particles with mass similar to that of the electron and whose kinetic energies extend into the keV
energy range.\footnote{An alternative possibility 
involving (tightly bound) electron scattering with GeV scale dark matter particles might also be possible \cite{dama44},
although it is significantly constrained \cite{kopp}.} 
Dark matter particles with these properties 
could then produce only 
observable (keV) electron recoils since keV nuclear recoils would then be kinematically suppressed.
To explain the DAMA annual modulation signal requires that the
flux of these particles in the relevant energy range (few keV)  annually modulates
(at least a few percent) with phase $\sim$ June $1^{st}$.

We will argue in the present paper, that 
kinetically mixed mirror dark matter is an example of a theory with the right broad features to potentially explain the DAMA annual modulation
signal consistently via DM - electron scattering. Similar, but more generic hidden sector models with light MeV scale dark matter particles
are, of course, also possible.
Recall,
mirror dark matter presumes the existence of a hidden sector which is exactly isomorphic to the standard model.
That is, the fundamental interactions are described by the Lagrangian \cite{flv}:
\begin{eqnarray}
{\cal L} = {\cal L}_{SM} (e, \mu, u, d, A_\mu, ...) +
{\cal L}_{SM} (e', \mu', u', d', A'_\mu, ...) + {\cal L}_{mix} \ .
\label{1}
\end{eqnarray}
The theory contains an exact parity symmetry: $x \to -x$ 
provided that left and right handed chiral fields are swapped in the mirror sector. 
Importantly, this parity symmetry is assumed to be unbroken 
by the vacuum; that is the Higgs and mirror Higgs doublets have exactly the same vacuum structure:
$\langle \phi \rangle = \langle \phi' \rangle$ \cite{flv}.
This means that for each type of ordinary particle there is an exactly degenerate mirror 
partner. That is, dark matter consists of a spectrum of dark matter particles of known masses: $e', H', He', O', ...$ \ .
It will be argued in the present paper that the scattering of the 
mirror electron component, with mass $m_e = 511.0$ keV, might be responsible for the DAMA annual modulation signal.

The ${\cal L}_{mix}$ 
part in Eq.(\ref{1}) describes possible interactions coupling the two sectors together. 
As in previous works, we consider
the kinetic mixing of the $U(1)_Y$  and $U(1)'_Y$ gauge bosons - a gauge invariant and renormalizable
interaction \cite{he} - which implies also  
photon - mirror photon kinetic mixing:
\begin{eqnarray}
{\cal L}_{mix} = \frac{\epsilon}{2} F^{\mu \nu} F'_{\mu \nu}
\ .
\label{kine}
\end{eqnarray}
Here $F_{\mu \nu}$ and $F'_{\mu \nu}$ denote the field strength tensors for the photon
and mirror photon respectively. 
The kinetic mixing interaction  
gives the mirror electron and mirror proton a 
tiny ordinary electric charge, $\epsilon e$ \cite{holdom}. 
This enables mirror nuclei to scatter off ordinary nuclei and
mirror electrons to scatter off ordinary electrons (essentially Rutherford
scattering if the electrons are free). 

%\vskip 0.5cm
\section{Cosmology with mirror dark matter}
%\vskip 0.2cm

The cosmology and astrophysics of such kinetically mixed mirror dark matter 
is somewhat nontrivial, see e.g. \cite{ber,ign,sph,sil,fc,paolo2,foot69,fstudy},
and also the recent review \cite{mmreview} (and references therein for a more extensive
bibliography).
The outcome of this work is that mirror dark matter can be the inferred
dark matter in the Universe provided kinetic mixing exists with strength $\epsilon \sim 10^{-9}$.
In this scenario,
dark matter halos in spiral galaxies are 
(currently) composed of mirror particles 
in a pressure supported multi-component plasma containing $e'$, $H'$, $He'$, $O'$, $Fe'$,...\cite{sph}.  
Such a plasma would dissipate energy
via thermal bremsstrahlung and other processes and so an energy source is needed to stabilize the halo.
Several studies \cite{sph,fstudy} have found that ordinary type II supernovae can supply 
the required energy if photon - mirror photon kinetic mixing has strength $\epsilon \sim 10^{-9}$ 
and the halo contains a significant mirror metal component ($\stackrel{>}{\sim} 1\%$ by mass).

Mirror particle self interactions act to keep the various components of the mirror particle plasma
in thermal equilibrium with a common temperature, $T$.
The temperature of the halo plasma, at the Earth's location, $T$ can be
roughly estimated by assuming 
hydrostatic equilibrium \cite{sph}:
\begin{eqnarray}
T \simeq \frac{1}{2} \bar m v_{rot}^2 
\ .
\label{rotx}
\end{eqnarray}
Here $v_{rot} \sim 240$ km/s is the Milky Way's galactic rotational velocity and
$\bar m = \sum n_{A'} m_{A'}/\sum n_{A'}$ is the mean mass of the
halo mirror particles (the sum includes the $e'$ component as the temperature is such
that mirror dark matter is typically ionized within spiral galaxies). 
Early Universe cosmology suggests that mirror helium is the dominant nuclear 
component \cite{ber}, with mirror BBN computations indicating\footnote{We adopt natural units $\hbar = c = 1$ unless otherwise stated.}  
$\bar m \approx 1.1 $ GeV 
for $\epsilon \sim 10^{-9}$ \cite{paolo2}.  
There are, of course, significant uncertainties, possibly around 20-30\%, in the halo temperature $T$
(which can be modelled by considering a similar variation in $\bar m$).
\footnote{
The analytic estimate of Eq.(\ref{rotx}) assumed an isothermal
halo. More realistically, the temperature is not expected to be spatially constant, but increases
towards the galactic center. Nevertheless, the numerical work of \cite{fstudy} suggests that the
temperature at the Sun's location is consistent with the estimate of Eq.(\ref{rotx}) to within around 20-30\%.} 

In a reference frame where there is no bulk halo motion,
the halo velocity distribution should be Maxwellian and thus $f_{i} = e^{-E/T}$
[$i$ denotes the type of mirror particle, $i = e', H', He', O', ...$].
The halo
particles are nonrelativistic, which means that $E = m_{i} |{\bf{u}}|^2/2$. It follows that 
the halo velocity distribution has the 
general form: 
\begin{eqnarray}
f_{i} = e^ {-|{\bf{u}}|^2/v_0^2
}
\end{eqnarray}
where 
\begin{eqnarray}
v_0(i) &=& \sqrt{{2T \over m_{i}}} 
        \simeq  v_{rot} \sqrt{{\bar m \over m_{i}}}\ . 
\label{v0}
\end{eqnarray} 
Observe that
the quantity $v_0 (i)$, which characterizes the velocity dispersion of the particle $i$,
depends on the mass of the particle. 
With $\bar m = 1.1$ GeV,
Eq.(\ref{v0}) indicates that
mirror electrons have velocity dispersion $v_0 (e') \approx 10,000$ km/s
while mirror helium ions have $v_0 (He') \approx 100$ km/s.

%\vskip 0.7cm
\section{Effects of mirror electromagnetic fields}
%\vskip 0.2cm

Halo mirror electrons can interact with the atomic electrons in a detector, 
providing them with $\sim$ keV recoils.
These recoils can potentially be detected in experiments, especially those which don't 
discriminate against electron recoils such as DAMA and 
CoGeNT. Previous
estimates have found that the average rate in which mirror electrons scatter off bound atomic electrons 
can potentially produce keV recoils at an observable
rate ($\stackrel{>}{\sim} 1$ cpd/kg/keV) for $\epsilon \sim 10^{-9}$ \cite{footelec,foot2}.

Subsequent to \cite{footelec,foot2} it was realized that the naive flux estimates of mirror electrons
at the Earth's location must be modified due to the influence of mirror electromagnetic fields produced
by captured mirror particles within the Earth \cite{mmreview}. Such effects most strongly influence the mirror electrons
because these particles are much lighter than the mirror ions.
Although these effects may be difficult to precisely estimate,
they are necessarily important (as we will discuss shortly).
In the absence of mirror electric (${\bf E'}$) or magnetic fields (${\bf B'}$),
the rate at which mirror electrons arrive at the Earth's surface (from outside the Earth) 
is approximately 
\begin{eqnarray}
R_{e'} &=&  4\pi R_E^2 n_{e'} \ \langle |v_{e'}^z| \rangle/2 
\nonumber \\
&\approx & 2\sqrt{\pi} R_E^2 n_{e'} v_0 (e') \ \ \ {\rm for} \ \ {\bf E'} = {\bf B'} = 0
\label{fluxe}
\end{eqnarray}
where $R_E$ is the Earth's radius and $v^z_{e'}$ is the component of the velocity 
normal to the Earth's surface.
In the second equation, above, we have used
$\langle |v_{e'}^z| \rangle \approx v_0 (e')/\sqrt{\pi}$, which is valid given that
the mirror electron velocity dispersion is much greater than the
Earth's speed through the halo. 

In the absence of mirror electric (${\bf E'}$) or magnetic fields (${\bf B'}$),
the corresponding (average) rate at which mirror nuclei (taken here to consist solely of mirror helium, $He'$)
arrive at the Earth's surface is roughly: 
\begin{eqnarray}
R_{He'} \approx \pi R_E^2 n_{He'} \ \langle v_{E} \rangle
\end{eqnarray}
which is valid in the limit where the mirror helium velocity dispersion, $v_0 (He')$, is much less than the Earth's average speed
through the halo: $\langle v_E \rangle \approx v_{rot} \sim 240$ km/s
[from Eq.(\ref{v0}), $v_0 (He')/v_{rot} \approx 0.5 $ for $\bar m = 1.1$ GeV]. 
Thus, in the absence of mirror electromagnetic fields,
the mirror electron flux arriving at the Earth is larger than the mirror nuclei flux by a factor of around
\begin{eqnarray}
{R_{e'} \over R_{He'}} &\approx &
{4v_0(e') \over \sqrt{\pi} v_{rot}} 
\nonumber \\
&\approx & {4 \over \sqrt{\pi}} \sqrt{{\bar m \over m_e}} \approx 100\ .
\end{eqnarray}

In reality, of course, this could not be the case. 
A greater mirror electron flux would lead to a larger mirror electron capture rate in
the Earth cf. the capture of mirror nuclei. 
This would lead to a rapidly increasing 
mirror electric charge within the Earth, $Q'_E e$.
(Mirror electrons are captured by collisions with captured mirror atoms \cite{fdiurnal} and potentially
also with ordinary atoms which can be important for $\epsilon \sim 10^{-9}$ cf. \cite{foot04}.)
In fact, in the absence of mirror electromagnetic fields
the expected capture rate, $R_{e'}^{C}$, can be estimated by slightly modifying Eq.(\ref{fluxe}): 
\begin{eqnarray}
R_{e'}^{C} &\approx & 2\sqrt{\pi} R_0^2 n_{e'} v_0 (e') \nonumber \\ 
       & \sim & 10^{26} \left({R_0 \over 4000 \ {\rm km/s}}\right)^2 \ {\rm s}^{-1} \ \ \ \ {\rm for} \ \ {\bf E'} = {\bf B'} = 0
\end{eqnarray}
where the capture radius, $R_0 \approx 4000$ km, has been deduced in ref.\cite{fdiurnal}.
This means that on time scales less than a second, a net mirror electric charge in the Earth
would be generated which would prevent any mirror electrons from reaching the Earth
(the Coulomb barrier exceeds the kinetic energy $Q'_E \alpha/(4\pi R_E) \gg T$).

Presumably,  mirror electric and magnetic fields would be generated such that
the flux of mirror electrons is reduced until it approximately 
matches the flux of mirror nuclei arriving at the Earth's surface.
That is, considering just the predominant $e', \ He'$ halo components, we expect that the 
actual rates averaged over the Earth's surface to roughly satisfy 
$R_{e'} \simeq 2R_{He'}$ (the factor of two is due to the charge ratio). 
As the flux of mirror nuclei varies during the year, due to the $\sim 7\%$ variation
in $v_E$,
one expects the flux of mirror electrons to correspondingly vary by the same amount.
[A larger modulation is possible if the mirror electromagnetic fields significantly suppress also the 
mirror ion capture rate in the Earth.]
If this happens, then it will be very important. It means that the rate at which mirror electrons scatter off target
electrons should annually modulate
with a sizable amplitude of at least $\sim 7\%$ and an expected phase of June $1^{st}$.

Additionally, one also expects the mirror electron flux to vary at different locations on the Earth's surface, and at the same
location at different times due to the Earth's daily rotation.
This is because the induced ${\bf E'}$ and ${\bf B'}$ fields 
are not expected to be spherically symmetric.
The mirror helium ions arriving at the Earth's surface come, predominately, from a fixed direction 
and are stopped in the Earth in a particular region, which we here denote as $Q$, significantly offset from the Earth's center. 
For an observer on the Earth the
position of this region changes during the day due to the Earth's rotation.
Naturally, the induced ${\bf E'}$ and ${\bf B'}$ fields 
at the detector's vicinity
should depend on the detector's position relative to this deposited charge and should therefore diurnally modulate. 
In fact, one might suspect that the effect of the induced ${\bf E'}$ and ${\bf B'}$ fields
should be weakest, and hence the mirror electron flux largest, when the detector is located at the greatest distance
from the region, $Q$.
If this reasoning is correct, then this tells us the phase of the expected diurnal modulation.
Of course, it is difficult to estimate the amplitude of the diurnal variation, without 
detailed modelling of the charge flows in and around the Earth, which is beyond the scope of this initial study.

Recently, the DAMA collaboration have published their results binned in sidereal hours \cite{damadiurnal}.
Figure 2 of \cite{damadiurnal} 
shows the diurnal residual rate, with average rate measured to be around 1 cpd/kg/keV.
The expected maximum rate of mirror electron scattering off ordinary electrons occurs at $T=8$ (sidereal) hours and
minimum at $T=20$ hours, given the above reasoning and DAMA's $T=0$ convention.
Considering the
$2-4$ keV energy range (i.e. the lowest currently available energy range 
where the electron scattering signal should be largest)
we can divide the data into two 12 hour bins: $B1$ for $T=8\pm 6$ hours and $B2$ for $T=20\pm 6$ hours.
The measured ratio $R\equiv B1/B2$ is then:
\begin{eqnarray}
R(measured) = 1.0072\pm 0.0031.
\end{eqnarray}
That is, $R(measured)$ is different from $1.0$ at approximately 2.3$\sigma$ C.L.
We have assumed only statistical errors, which seems reasonable given that there should not be any significant systematic effects 
related to sidereal time. This appears to be an interesting hint, which can obviously be further
checked as more data are accumulated. In particular, the forthcoming DAMA/Libra results with lower energy threshold
might be particularly interesting as the electron scattering rate
is expected to be larger at lower energies.
To make detailed predictions for the annual and diurnal modulation spectrum 
is difficult without 
accurate modelling of the ${\bf E'}$ and ${\bf B'}$ induced in the Earth.
In the following, we will use a highly simplified model, and focus only on the annual
modulation component.

Since the high velocity part of the mirror electron spectrum should be least affected by mirror electromagnetic fields,
it might be possible to model
the mirror electron velocity spectrum at the Earth with a cutoff, $v_c$:
\begin{eqnarray}
f_{e'} &=& e^{-|{\bf{v}}|^2/v_0^2} 
\ \ {\rm for}\ \ |{\bf{v}}| > v_c \nonumber \\
f_{e'} &=& 0 \ \ {\rm for}\ \ |{\bf{v}}| < v_c 
\ .
\label{cutt}
\end{eqnarray}
We further make the simplifying assumption that the mirror ion flux is approximately unaffected by mirror electromagnetic fields.
Although rough estimates of the expected ${\bf E'}$ and ${\bf B'}$ fields, to be presented elsewhere, suggest that this
is a reasonable assumption, a detailed study is needed. A reduction in the ion flux, even a moderate one, would
be quite important, and would likely increase the fractional modulation ($\stackrel{>}{\sim} 7\%$). 

The cutoff $v_c (t)$ can be determined by imposing $R_{e'} \simeq 2R_{He'}$.
Again considering just the predominant $e', \ He'$ components, 
this condition implies a relation for $v_c$:
\begin{eqnarray}
\int^{\infty}_{|{\bf{v}}| = v_c} {e^{-[|{\bf{v}}|/v_0(e')]^2} \over [v_0(e')]^3}\ |v^z| d^3 v \simeq 
{v_E \over 2} \int_0^{\infty} {e^{-[|{\bf{v}}|/v_0(He')]^2} \over [v_0(He')]^3}
\ d^3 v
\end{eqnarray}
or
\begin{eqnarray}
{v_E(t) \over v_0(e')} \simeq {2 \over \sqrt{\pi}} \ e^{-[v_c/v_0(e')]^2} \ \left(1 + \left[ {v_c\over v_0(e')}\right]^2\right) 
\ .
\label{mod8}
\end{eqnarray}
The temporal variation of the mirror electron flux at the Earth's surface is encoded in the time dependence of the cutoff, $v_c (t)$, which 
annually modulates with time due to the modulation of $v_E (t)$:
\begin{eqnarray}
v_E (t) &=& v_{\odot} + v_{\oplus} \cos\gamma \cos \omega (t - t_0)
\nonumber \\
&=& v_{\odot} + \Delta v_E \cos \omega (t - t_0)
\ .
\end{eqnarray}
Here $v_{\odot} = v_{rot} + 12 \ {\rm km/s}$ is the Sun's speed
with respect to the galactic halo
and $v_{\oplus} \simeq 30$ km/s is the Earth's orbital
speed around the Sun. The phase is $t_0 = 152.5$ days and 
$\omega = 2\pi/{\rm year}$.
The angle $\gamma \simeq 60^o$ is the inclination of the Earth's orbital plane relative
to the galactic plane and $\Delta v_E \simeq 15$ km/s.

\vskip 0.7cm
\section{The DM-electron scattering rate, $dR_{e}/dE_R$}
\vskip 0.1cm

We now discuss the interaction cross section describing the scattering of mirror
electrons off ordinary electrons.
Photon - mirror photon kinetic mixing induces a small
coupling between ordinary photons and mirror electrons, of magnitude $\epsilon e$.
This enables mirror electrons to scatter off ordinary electrons, taken here
to be free, with
cross section:
\begin{eqnarray}
{d\sigma \over dE_R} = {\lambda \over E_R^2 v^2}
\label{cs}
\end{eqnarray}
where 
\begin{eqnarray}
\lambda \equiv {2\pi \epsilon^2 \alpha^2 \over m_e} \ .
\end{eqnarray}
Here $E_R$ is the recoil energy of the target electron, initially presumed at
rest relative to the incoming mirror electron  of velocity $v \ll c$.
This cross section should also approximate the scattering of halo mirror electrons off
bound atomic electrons provided that
the recoil energies are much larger than the binding energy.
For an NaI pair, there are 54 electrons with binding energy less than around 1 keV.  
In the crude analysis to follow, we treat these 54 electrons as free, and ignore the
interactions of the 10 most tightly bound electrons (with binding energy $\stackrel{>}{\sim}$ 1 keV).
The point of this crude analysis is to investigate whether DM-electron scattering has the potential to 
explain the DAMA annual modulation signal.  Naturally a more sophisticated analysis could be done, and is clearly warranted, to
more exactly model the cross section.

With our simplifying approximations, the predicted differential interaction rate is:
\begin{eqnarray}
{dR_e \over dE_R} &=& 
gN_T  n_{e'} \int {d\sigma \over dE_R}
{f_{e'}(v) \over k} |{\bf{v}}|
d^3v \nonumber \\
&=& g N_T  n_{e'}
{\lambda \over E_R^2 } \int^{\infty}_{|{\bf{v}}| > v_{min}
(E_R)} {f_{e'}(v) \over k|{\bf{v}}|} d^3 v 
\label{55}
\end{eqnarray}
where $N_T$ is the number of target NaI pairs per kg of detector,  
$k = [\pi v_0^2 (e')]^{3/2}$ is the Maxwellian distribution normalization factor
and $n_{e'}$ is the halo mirror electron number density.
Using the standard value of $\rho_{dm} = 0.3\ {\rm GeV/cm}^3$ for the dark matter mass density 
and assuming that the halo mass is dominated by the $H', He'$ component [with an $e', H', He'$ mean mass
of $\bar 1.1$ GeV] , we find that $n_{e'} \simeq 0.17\ {\rm cm}^{-3}$. 
The quantity $g = 54$ is the number of loosely bound atomic electrons in each NaI pair
as discussed above. 
Also, the lower velocity limit,
$v_{min} (E_R)$, 
is given by the kinematic relation:
\begin{eqnarray}
v_{min} &=& \sqrt{ {2E_R\over m_e } } \ .
\label{v}
\end{eqnarray}
With the distribution given by Eq.(\ref{cutt}), the
velocity integral in Eq.(\ref{55}) can be analytically solved leading to:
\begin{eqnarray}
{dR_e \over dE_R} &=& 
g N_T n_{e'} {\lambda \over E_R^2}
\ {2e^{-x^2}
\over \sqrt{\pi} v_0 (e') }
\label{r68}
\end{eqnarray}
where $x = MAX[v_{min}/v_0, v_c/v_0]$.
Observe that the rate modulates in time ($\sim 7\%$) at low
recoil energy ($E_R \stackrel{<}{\sim} m_e v_c^2/2$) due to the modulation of $v_c$
[obtained by numerically solving Eq.(\ref{mod8})].

We are now almost ready to estimate the DM-electron scattering 
rate for DAMA.
In order to compare with the experimentally measured rate, we need to
convolve the rate, with a Gaussian to take into account the detector
resolution:
\begin{eqnarray}
{dR_e \over dE_R^m} &=& 
{1 \over \sigma \sqrt{2\pi}} 
\int {dR_e \over dE_R} e^{-(E_R - E_R^m)^2/2\sigma^2}  \ dE_R
\ .
\label{bla}
\end{eqnarray}
Here $E_R^m$ is the `measured recoil energy' while the actual recoil energy is 
denoted as $E_R$, and $\sigma$ describes the resolution. 
[We use $\sigma/E_R = 0.448/\sqrt{E_R (keV)} + 0.0091$, which is
the central value measured by the DAMA collaboration \cite{dama66}.]

In figure 1a we present results for the DAMA annual modulation amplitude for
a representative choice of parameters. As the figure shows, for $\epsilon \sim 10^{-9}$
the total modulation above $E_R > 2$ keV is of the right magnitude of interest;
however the predicted
spectrum is significantly steeper than the data. This might well be a result of the
simplified approximations we have used. Certainly, a more realistic modelling of the
scattering cross section of DM off the bound atomic electrons would be useful (including also the 10 most tightly bound
electrons), and also
a more sophisticated modelling of the $e'$ velocity distribution beyond the simple cutoff
prescription of Eq.(\ref{cutt}).
In figure 1b we give the results for the unmodulated (average) DM - electron scattering rate 
anticipated for DAMA for the same parameters as per figure 1a. 

Similar results, for both modulated and unmodulated rates, are
expected for other experiments, and are within current limits, e.g. \cite{cdmslowelectron,cogent,xmas,cdex}.
A possible
exception is the double phase nobel liquid experiments such as LUX \cite{lux} and XENON100 \cite{xenon}.
However this latter class of experiments utilizes large electric fields which can induce mirror electron
charge flows and hence mirror magnetic fields with strength proportional to the density of captured mirror dark matter (more precisely
the free mirror electron density at the detector's location).
Rough estimates indicate that a mirror electron number density at the detector's location of around $10^{12} \ {\rm cm^{-3}}$ 
is sufficient to induce large enough $B'$ fields to shield the LUX detector from halo mirror electrons (with $\epsilon \sim 10^{-9}$ assumed).
Such a density is quite substantial, but nevertheless appears to be possible.
If this is the physical effect responsible for suppression of the rates in LUX and XENON100, then placing the DAMA
detector in (or near) a large electric field should also suppress the rate for that experiment. Comparison
of the rates with field on and off could then be used to check this explanation.

If DM - electron scattering is responsible for the annual modulation signal observed
by DAMA, then
this can be probed by many other experiments. In particular, the large xenon experiments such
as LUX and XENON1T can search for an annual modulation in electron recoils (although as mentioned above, the
unmodulated rate might be suppressed as a result of the large electric field employed in that detector setup). Usually
such electron recoil events are not analysed but instead could be examined for an annual modulation.  
The single phase XMAS xenon experiment \cite{xmas} does not employ any electric field
and should therefore be particularly sensitive to this electron scattering
interpretation of the DAMA signal. Analysis of around a year's data from this experiment
is expected shortly.
Also, low threshold experiments such as
CoGeNT \cite{cogent}, CDEX \cite{cdex}, and C4 \cite{c4} can also probe the DM - electron scattering signal.
In fact, it might be possible to explain
the modulation observed by CoGeNT at low recoil energies \cite{cogent2} 
within this DM - electron scattering framework.

\vskip 1.2cm

\centerline{\epsfig{file=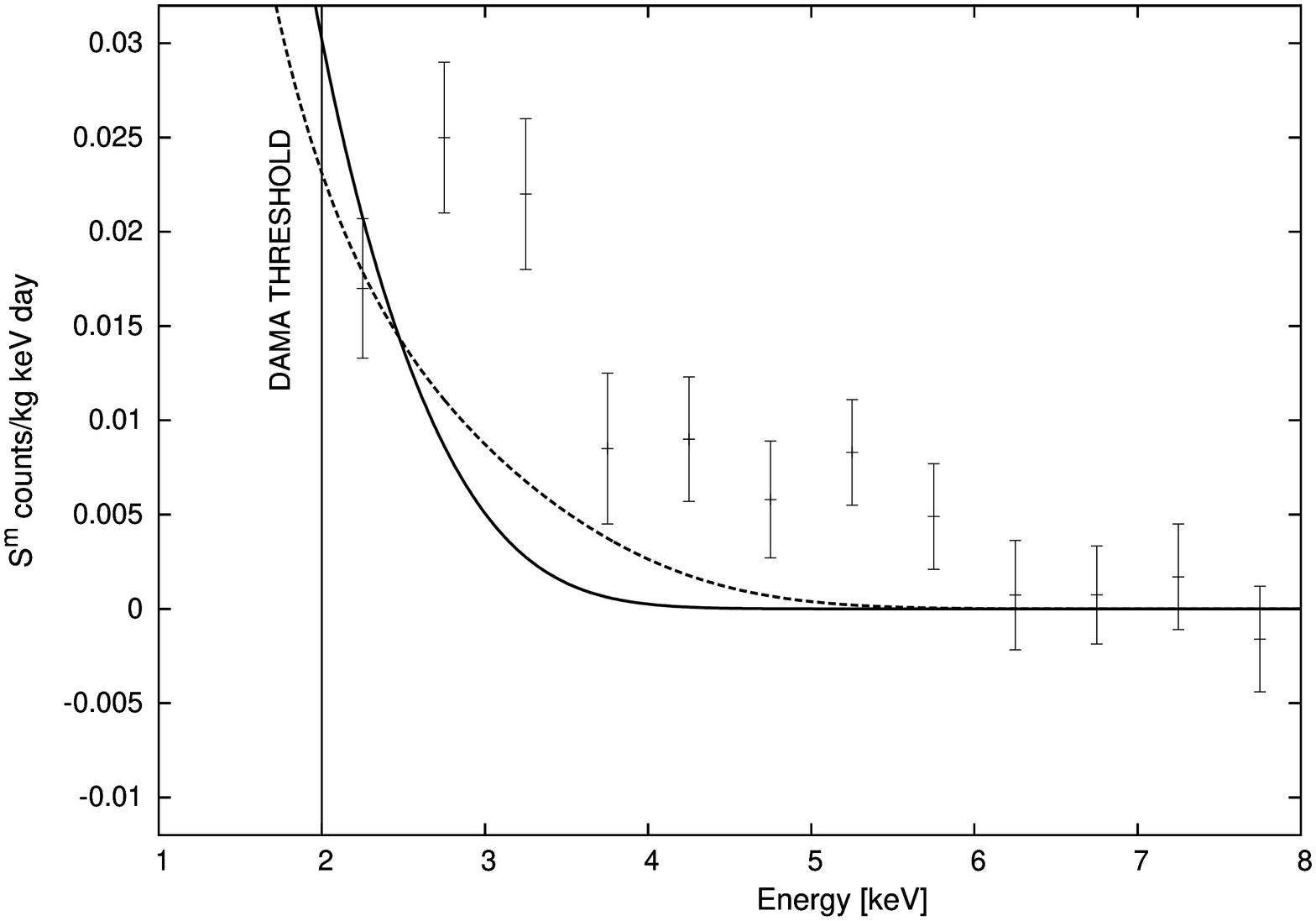,angle=270,width=12cm}}
\vskip 0.5cm

\noindent
{\small Figure 1a: DAMA modulation amplitude, $S^m$, due to DM - electron scattering in the simplified model discussed in the text.
The solid line corresponds to standard halo mean particle mass value of $\bar m = 1.1$ GeV, while the dashed line is for $\bar m = 1.6$ GeV.
Both curves assume $v_{rot} = 260$ km/s and $\epsilon = 10^{-9}$.}
\vskip 0.3cm

\centerline{\epsfig{file=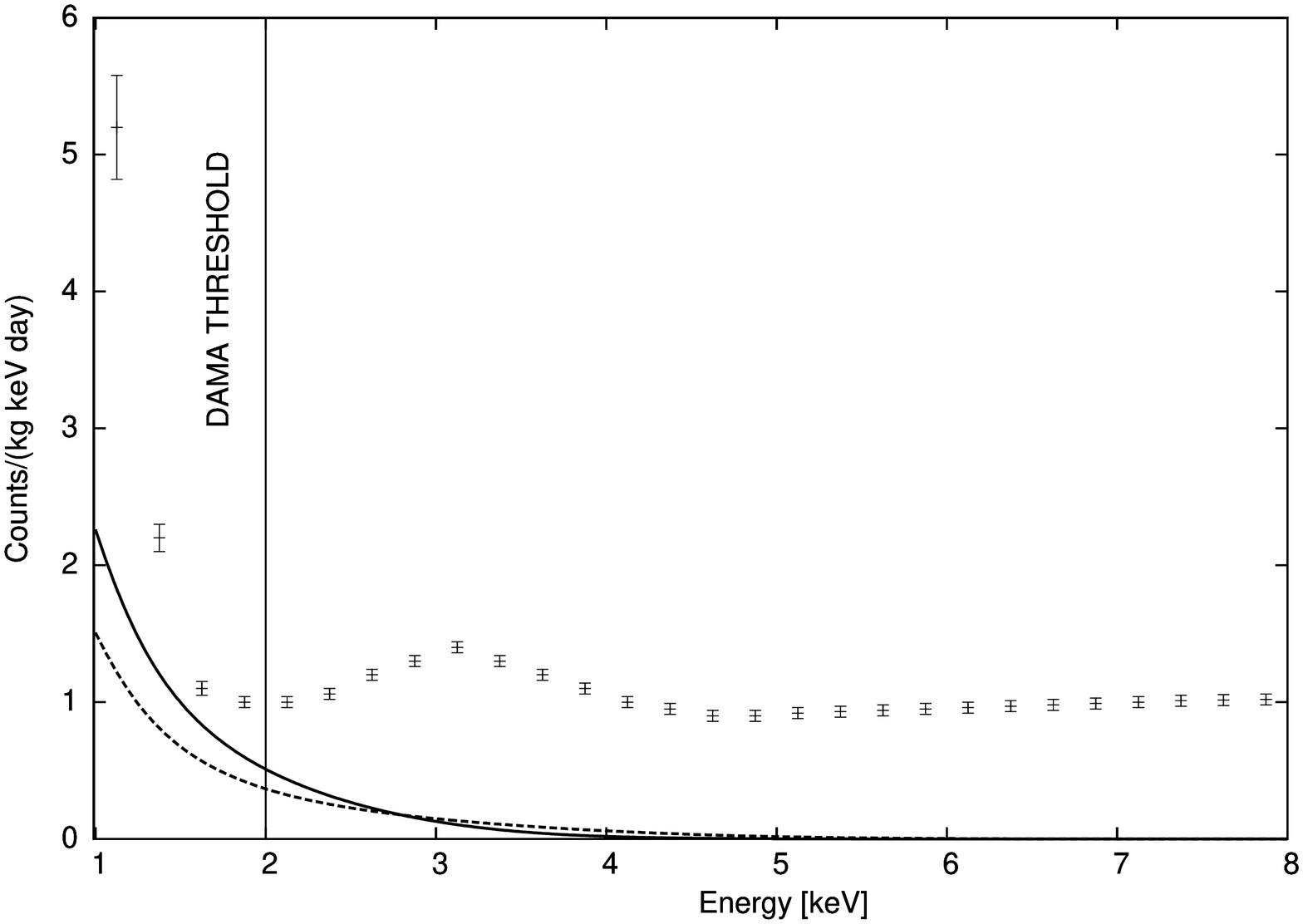,angle=270,width=12cm}}
\vskip 0.5cm

\noindent
{\small Figure 1b: DAMA average rate due to DM - electron scattering in the simplified model discussed in the text for the same 
parameters as per figure 1a.}

\vskip 1.5cm

Finally, let us note that within the context of kinetically mixed mirror dark matter, mirror ion scattering
off ordinary nuclei is also expected and has been discussed
in a number of publications \cite{footnew} (and references there-in). While
mirror helium is too light to produce observable recoils in current experiments, mirror metal components can
potentially be detected.
Indeed astrophysical arguments suggest that a substantial
mirror metal component is needed with mass fraction, $\xi_{A'}$, greater than around 1\% \cite{fstudy}.
Considering (as an example) mirror oxygen as the dominant mirror metal component,
a $1\%$ mirror metal mass fraction indicates $\epsilon \sqrt{\xi_{O'}}$ of around $10^{-10}$ (if $\epsilon \sim 10^{-9}$).
Such parameters are consistent \cite{footsoon,mmreview} with the null results reported by LUX \cite{lux}, XENON100 \cite{xenon}, SuperCDMS \cite{cdms},
CRESST-II \cite{cresst} etc  and will be probed by current and future experiments.

\section{Conclusion}

To conclude, we have considered an explanation of the DAMA  
annual modulation signal in terms of dark matter - electron scattering.
Such an explanation is now favoured given the stringent constraints on nuclear recoil rates obtained by LUX, SuperCDMS
and other experiments. 
Multi-component dark matter models featuring light MeV scale
dark matter particles can potentially explain  
the experiments via electron scattering.
We have focussed attention on a specific such theory, kinetically mixed mirror dark matter,
which appears to have the
right broad properties to consistently explain the experiments.
If this is the explanation of the annual modulation signal found in the DAMA experiments
then a diurnal modulation signal (i.e. period of a sidereal day) is also anticipated.
We have pointed out that the data from the DAMA experiments 
show a diurnal variation at around 2.3$\sigma$ C.L. with phase consistent with that expected.
Importantly, this DM - electron scattering interpretation of DAMA can be tested in 
further results from the DAMA experiment, in
low threshold experiments such as CoGeNT, CDEX, C4 and 
potentially also in the large
XENON experiments (LUX, XENON1T,...) by searching for annually and diurnally modulated electron recoils.

\vskip 1cm
\noindent 
{\large \bf Acknowledgments}

\vskip 0.2cm
\noindent
This work was supported by the Australian Research Council. The author would like to thank R. Gaitskell for useful 
correspondence.

\end{document}